\documentclass[12pt]{article}

\newtheorem{theorem}{Theorem}
\newtheorem{lemma}{Lemma}

\newtheorem{corollary}{Corollary}

\newtheorem{remark}{Remark}

\begin{document}

\title{An Effective Upperbound on Treewidth Using Partial Fill-in of Separators}

\author{
Boi Faltings
\thanks{Artificial Intelligence Laboratory (LIA), Ecole Polytechnique F\'{e}d\'{e}rale de Lausanne (EPFL), 1015 Lausanne, Switzerland. E-mail: boi.faltings@epfl.ch }
\and
\addtocounter{footnote}{1}
Martin Charles Golumbic
\thanks{Caesarea Rothschild Institute and
Department of Computer Science, University of Haifa,
Mt. Carmel, Haifa 31905, Israel. E-mail: golumbic@cs.haifa.ac.il}
} 	
\date{\today}
\maketitle

\begin{abstract}

Partitioning a graph using graph separators, and particularly clique separators,
are well-known techniques to decompose a graph into smaller units which can be
treated independently.  It was previously known that the treewidth was bounded above
by the sum of the size of the separator plus the treewidth of disjoint components,
and this was obtained by the heuristic of filling in all edges of the separator
making it into a clique.

In this paper, we present a new, tighter upper bound on the treewidth of a graph obtained
by only partially filling in the edges of a separator. In particular, the method
completes just those pairs of separator vertices that are adjacent to a common component,
and indicates a more effective heuristic than filling in the entire separator.

We discuss the relevance of this result for combinatorial algorithms and give
an example of how the tighter bound can be exploited in the domain of constraint
satisfaction problems.

\end{abstract}


\maketitle


\section{Introduction}

Let $G = (V,E)$ be an undirected graph.
We denote by $G_X = (X,E_X)$  the subgraph of $G$ induced by $X \subseteq V$
               where $E_X = \{ (u,v) \in E \ | \  u,v \in X \}$.

A {\em tree decomposition} for a graph $G$ is defined as a tree $T$
whose nodes are labelled by subsets of $V$ called ``clusters'' (or ``bags'')
such that

	(1) every vertex $v \in V$ appears in at least one cluster,	

	(2) if $(u,v) \in E$, then $u$ and $v$ co-occur in some cluster,
and

	(3) for every $v \in V$, the set of nodes of $T$ which include $v$
in their cluster induces a connected subgraph (i.e., a subtree) of $T$,
denoted $T(v)$.

The {\em width} of a tree decomposition $T$ is the size of the largest
cluster minus 1, and is denoted by $width(T)$.

A given graph $G$ may have many possible tree decompositions, including
the trival representation as a single node with cluster equal to $V$.
The {\em treewidth} $tw(G)$ of a graph $G$ is defined to be the minimum
width over all tree decompositions for $G$.
Such a tree decomposition is called a {\em minimum tree decomposition} for $G$.

\begin{remark}
The treewidth of a tree equals 1, of a chordless cycle equals 2,
of a clique on $k$ vertices equals $k-1$, and of a stable (independent) set
equals zero.  It is also well known, that a chordal graph has a
minimum tree decomposition where each cluster is a maximal clique of the graph,
thus, the treewidth of a chordal graph is the size of its largest clique minus 1.
\end{remark}

The theory of treewidth, introduced by Robertson and Seymour \cite{RobS86},
is a very rich topic in discrete mathematics,
and has important algorithmic significance, since many NP-complete problems
may be solved efficiently on graphs with bounded treewidth.
The reader is referred to \cite{Bod93,Bod06,Klo94} for further treatment of the subject.

Partitioning a graph using graph separators, and particularly clique separators,
is well-known as a method to decompose a graph into smaller components which can be treated independently \cite{BodKos09}.
A previously known bound~(\cite{Dechter}) for the treewidth of a graph $G = (V,E)$ was based on identifying a separator $S \subset V$ such that
the graph $G_{V \backslash S}$ obtained by deleting from $G$ all vertices
in $S$ and their incident edges is broken into components $G_1,...,G_k$:
\[
tw(G) \leq |S| + max_i[tw(G_i)]
\]
This was obtained by the heuristic of filling in all edges of the separator
making it into a clique, so we call it the {\em separator-as-clique bound}.
It is important not only for estimating the treewidth of a graph, but
decompositions that result in a low bound on treewidth give rise to
efficient algorithms for a variety of problems on graphs.

In Section \ref{section-our-result}, we present a new, tighter upper bound on the treewidth of a graph whose novelty is filling in fewer edges of the separator. Our method completes just those pairs of separator vertices that are adjacent to a common component, giving a lower treewidth of the augmented supergraph. We thus call it the {\em separator-as-components bound}.
This is followed by an example in Section \ref{section-example}
to illustrate our method.
In Section \ref{section-motivation}, we conclude by discussing
its application to solving constraint
satisfaction problems combining search with dynamic programming,
which was our motivation for having studied the question of improving the bounds on treewidth.


\section{Our result}
\label{section-our-result}

We first recall the Helly property which is satisfied by subtrees of a tree \cite{Gol80}.
By definition, if $(u,v) \in E$ then $T(u) \cap T(v) \neq \emptyset$.
The Helly property for trees states that {\em if a collection of subtrees of a tree
pairwise intersect, then the intersection of the entire collection is nonempty.}
This immediately implies the following well-known (folklore) observation
\cite{BodMoh93}, which will be used below.

\begin{lemma}
\label{remark-helly}
            Let $T$ be a tree decomposition for $G$.
If $C$ is a clique of $G$, then there is a cluster $X$ (labelling a node of $T$)
such that $C \subseteq X$.

\end{lemma}

Let $G = (V,E)$ be an undirected graph and let $S \subseteq V$ be a
subset of the vertices.  We consider the connected components
$G_1, \ldots , G_t $ of $G_{V \setminus S}$, i.e., the connected subgraphs obtained
from $G$ by deleting all vertices of $S$ and their incident edges.
We denote by $V_i$ the vertices of $G_i$, that is, $G_i = (V_i,E_{V_i})$.
Finally, let $S_i \subseteq S$ denote the subset consisting of all vertices of $S$
which have neighbors in $G_i$.


Define $(x,y)$ to be a \emph{fill-in} edge
if $(x,y) \notin E$ and $x,y \in S_i$ for some $i$,
and let $F$ be the set of all fill-in edges.
Define the graph  $H =(V,E')$  to be the supergraph of  $G$ , where
 	$E' = E  \cup F$.
In other words, an edge is filled in between $u,v \in S$ in $E'$ if there
is a path in $G$ from $u$ to $v$ using only intermediate vertices of some component $G_i$.
Thus, each $S_i$ becomes a clique in $H_S$, the subgraph of $H$ induced by $S$.

The following is our new result:

\begin{theorem}

$tw(G) \leq \max_i \{tw(H_S),  |S_i|+ \{tw(G_i)\} \}$

\end{theorem}
\label{theo:little-lemma}

{\bf Proof.}
           Let $T_S$ be a minimum tree decomposition for the subgraph $H_S$,
and let $T_i$ be a minimum tree decomposition for $G_i$.
We will now construct a tree decomposition $T$ for $G$.

            Since the set $S_i$ forms a clique in $H_S$, by Remark~\ref{remark-helly},
there is a cluster $X_i$ in $T_S$ containing $S_i$.
To form $T$, we augment the union of $T_S$ and all the $T_i$ by
(i) adding the members of $S_i$ to each cluster of $T_i$,
and (ii) adding a new tree edge from the node $x_i$ with label $X_i$ to an
arbitrary node $v_i$ of $T_i$, for each $i$.

We now show that $T$ is a tree decomposition for $H$ and thus also for $G$.
Condition (1) of the definition of tree decomposition is trivial, and
condition (3) is proven as follows:
Each $T(v)$ for $v \in V \setminus S$ remains unchanged and is
therefore a subtree of $T$. Also, each $T(x)$ for $x \in S$ is a subtree of $T$
since it consists of the union of its former subtree $T_S(x)$ and,
for each $i$ in which $x$ has neighbors in $G_i$, the entire tree $T_i$ along
with the new edge $(v_i , x_i)$ connecting $G_i$ with the node with label $X_i$.

We prove condition (2) in three cases.

{\bf Case 1}: $u,v \in V \setminus S$:
                   If $(u,v) \in E$, then $u$ and $v$ are in the same connected
component, say $G_j$, and they appear together in some cluster at a node of $T_j$.

{\bf Case 2}: $u \in V \setminus S$ and $ v \in S$:
                   If $(u,v) \in E$ where $u$ is in the component $G_j$, then $v \in S_j$
and they now appear together in some (in fact, in every) cluster of $T_j$ where $u$ appears.

{\bf Case 3}: $u,v \in S$:
                   If $(u,v) \in E$, then $(u,v) \in E'_S$, so $u$ and $v$ co-occur
in some cluster at a node of $T_S$, hence in $T$.

Thus, $T$ is a tree decomposition for $H$ and thus also for $G$.

It now remains to show that $w = width (T)$ is at most
$\max \{tw(H_S),  |S_i| + tw(G_i) | i=1,\ldots,t \}$.

We first observe that $tw(G_{V \setminus S}) = \max \{ tw(G_i) | i = 1,\ldots,t \}$,
since the $G_i$ are disjoint graphs.

Let $Y$ be the largest cluster in $T$, that is, $w=|Y|-1$.
If $Y$ is the label of a node in $T_S$, then $w=tw(H_S)$.
Otherwise, $Y$ is the new label of a node in $T_j$ for some component $G_j$,
that is, $Y= S_j  \cup B$ where $B$ is the largest (original) cluster in $T_j$,
and $tw(G_j)=|B|-1$.  Therefore,
\[
        w=|Y|-1=|S_j|+|B|-1 = |S_j| + tw(G_j)
\]
which proves the claim.
Q.E.D.

\begin{corollary}

$tw(G) \leq tw(H_S) + tw(G_{V \setminus S}) + 1 $

\end{corollary}

{\bf Proof.}
This follows since $|S_i| \leq |X_i| \leq tw(H_S) + 1 $ for all $i$ and
$tw(G_{V \setminus S}) = \max_i \{ tw(G_i) \}$.

\bigskip

\begin{remark}
Our result can be seen as a strengthening of the notion of safe separators \cite{BodKos06}
and of w-cliques \cite{Dechter} where these authors fill-in all pairs of
vertices in $S$ making it a clique, and giving the weaker upperbound
$tw(G) \leq |S| + {\max}_i \{tw(G_i)\} = |S|  + tw(G_{V \setminus S})$.
\end{remark}


\section{Example}
\label{section-example}


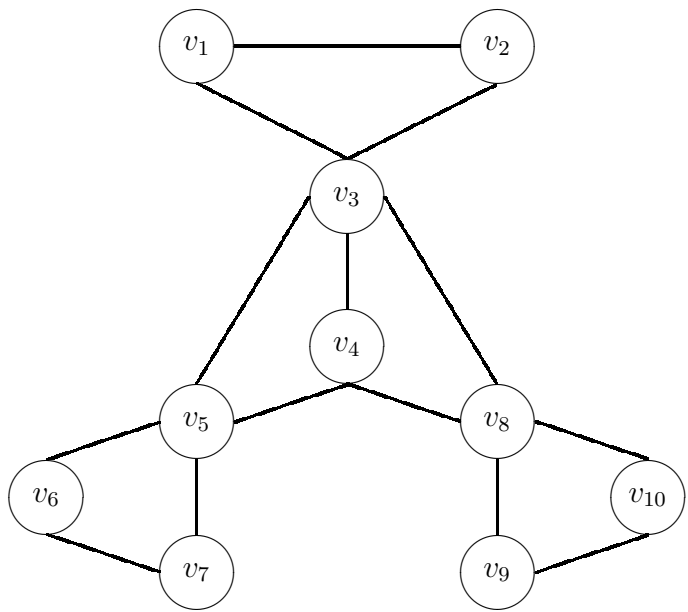
\begin{figure}

\ifx\JPicScale\undefined\def\JPicScale{1}\fi
\unitlength \JPicScale mm
\begin{picture}(110,90)(0,0)
\linethickness{0.3mm}
\put(45,85){\circle{10}}

\linethickness{0.3mm}
\put(65,65){\circle{10}}

\linethickness{0.3mm}
\put(85,85){\circle{10}}

\linethickness{0.3mm}
\put(50,85){\line(1,0){30}}
\linethickness{0.3mm}
\multiput(45,80)(0.24,-0.12){83}{\line(1,0){0.24}}
\linethickness{0.3mm}
\multiput(65,70)(0.24,0.12){83}{\line(1,0){0.24}}
\linethickness{0.3mm}
\put(65,45){\circle{10}}

\linethickness{0.3mm}
\put(45,35){\circle{10}}

\linethickness{0.3mm}
\put(85,35){\circle{10}}

\linethickness{0.3mm}
\put(25,25){\circle{10}}

\linethickness{0.3mm}
\put(45,15){\circle{10}}

\linethickness{0.3mm}
\put(105,25){\circle{10}}

\linethickness{0.3mm}
\put(85,15){\circle{10}}

\put(45,85){\makebox(0,0)[cc]{$v_1$}}

\put(85,85){\makebox(0,0)[cc]{$v_2$}}

\put(65,65){\makebox(0,0)[cc]{$v_3$}}

\put(65,45){\makebox(0,0)[cc]{$v_4$}}

\put(45,35){\makebox(0,0)[cc]{$v_5$}}

\put(25,25){\makebox(0,0)[cc]{$v_6$}}

\put(45,15){\makebox(0,0)[cc]{$v_7$}}

\put(85,35){\makebox(0,0)[cc]{$v_8$}}

\put(85,15){\makebox(0,0)[cc]{$v_9$}}

\put(105,25){\makebox(0,0)[cc]{$v_{10}$}}

\linethickness{0.3mm}
\put(65,50){\line(0,1){10}}
\linethickness{0.3mm}
\multiput(45,40)(0.12,0.2){125}{\line(0,1){0.2}}
\linethickness{0.3mm}
\multiput(50,35)(0.36,0.12){42}{\line(1,0){0.36}}
\linethickness{0.3mm}
\multiput(70,65)(0.12,-0.2){125}{\line(0,-1){0.2}}
\linethickness{0.3mm}
\multiput(65,40)(0.36,-0.12){42}{\line(1,0){0.36}}
\linethickness{0.3mm}
\multiput(25,30)(0.36,0.12){42}{\line(1,0){0.36}}
\linethickness{0.3mm}
\put(45,20){\line(0,1){10}}
\linethickness{0.3mm}
\put(85,20){\line(0,1){10}}
\linethickness{0.3mm}
\multiput(90,15)(0.36,0.12){42}{\line(1,0){0.36}}
\linethickness{0.3mm}
\multiput(25,20)(0.36,-0.12){42}{\line(1,0){0.36}}
\linethickness{0.3mm}
\multiput(90,35)(0.36,-0.12){42}{\line(1,0){0.36}}
\end{picture}

\caption{Example graph}
\label{fig:example}
\end{figure}


Consider the example graph shown in Figure~\ref{fig:example}.
It has a tree decomposition into the following cliques:
\begin{eqnarray*}
C_1 & = & \{v_1,v_2,v_3\} \\
C_2 & = & \{v_3,v_4,v_5\} \\
C_3 & = & \{v_3,v_4,v_8\} \\
C_4 & = & \{v_5,v_6,v_7\} \\
C_5 & = & \{v_8,v_9,v_{10}\}
\end{eqnarray*}
that are all of size 3, and the subgraph with vertices $\{v_1,v_2,v_3\}$
has no tree decomposition into smaller cliques. Thus, its treewidth is 2.
(In fact, it a chordal graph, and $C_1$--$C_5$ a clique decomposition.)

To illustrate our method, and provide an example for the application
described in Section \ref{section-motivation},
choose $S = \{v_3,v_4,v_5,v_8\}$, thus leaving three connected components
$G_1 = \{v_1,v_2\}$, $G_2 = \{v_6,v_7\}$ and $G_3 = \{v_9,v_{10}\}$.
Each of these has a treewidth of 1.

Using the separator-as-clique bound, we upper-bound the treewidth of G as:
\[
tw(G) \leq |S| + max_{i}\{tw(G_i)\} = 4 + 1 = 5
\]

Using Theorem~\ref{theo:little-lemma}, the separator-as-components method obtains a tighter bound, as follows.
Note that we have $S_1 = \{v_3\}$, $S_2 = \{v_5\}$ and $S_8 = \{v_8\}$,
that $H_S = G_S$ since none of the $G_i$ is connected via multiple
vertices, and that $tw(H_S) = 2$. Now we have:
\[
tw(G) \leq max_{i}\{tw(H_S), |S_i|+tw(G_i)\} =  max_{i}\{2, 1+1\} = 2
\]
which is exactly the treewidth of $G$.

To be fair, we should note that for the separator-as-clique bound, the best
possible choice for $S$ would have been $S' = \{v_3,v_5,v_8\}$, thus
leaving an additional disjoint component $G_4 = \{v_4\}$ and giving a bound
of $|S'|+1 = 4$ instead of $5$. Using our separator-as-component method, this separator would not give a bound that is as good because the $S_4$ of neighbours of the
new component $G_4$ includes all 3 vertices in $S'$, thus $|S_4|+tw(G_4) = 3+0 = 3$, and the bound will be 3. While this is still better than the separator-as-clique bound, it is
a counterintuitive indication that the smaller separator $S'$ does not give
the best decomposition. This fact is important in the application example below.


\section{Application to Constraint Satisfaction Problems}
\label{section-motivation}

Although this paper may be regarded as purely mathematical, it has its motivation in an important heuristic method for solving various problems
that are commonly solved using search algorithms, including constraint satisfaction~(\cite{PetFal07}), satisfiability and Bayesian inference~(\cite{Dechter}).

In search algorithms, there is a tradeoff between (1) the time complexity of
searching for a solution, (2) the size of the memory (or cache) to
store intermediate computations, and (3) for distributed implementations, the communication complexity for sending and sharing information between parts of the graph. Balancing these three parameters within the resources available
is the basis of our motivation.

As an example, consider a constraint satisfaction problem (CSP)~\cite{CP-handbook}), where each of a set of variables $X = \{x_1,...,x_n\}$ has to be
assigned a value in the corresponding domains $D = \{d_1,...,d_n\}$ such
that for each of a set of constraints $C = \{c_1,...,c_m\}$, the values
assigned to the variables in the scope of the constraint are among
its allowed tuples. When all constraints have a scope of one or two
variables, the CSP can be represented as a graph whose vertices are the
variables and whose edges are the constraints.

Constraint satisfaction problems are commonly solved using backtrack
search algorithms that assign values to the variables one at a time and
backtrack as soon as no value consistent with previous assignments can be
found. These have a complexity of $O(|d|^n)$.

However, the efficiency of search can often be significantly improved using
caching of partial results. In particular, when the constraint graph
has a small separator $S$ that splits the graph into at least two
components, one can record all value combinations
of variables in the separator that admit a consistent assignment to
the variables in one of the components and then replace backtrack
search through these variables by table lookup for the rest of the graph.
Equivalently, one can use dynamic programming to determine which
combinations of values for variables in the separator yield a
consistent assignment.
It has been shown~\cite{Dechter} that the time complexity of such an
algorithm is at least $O(|d|^{|S|+max_i[tw(G_i)]})$ and can be
much less than $O(|d|^n)$.

Let the example graph of Figure~\ref{fig:example} represent
a CSP with 10 variables with $d$ possible values each, where the arcs correspond to arbitrary unstructured constraints. Backtrack search would require
time on the order of $O(d^{10})$, but memory only linear in $d$.
However, search with caching or dynamic programming can solve this problem in cubic time and quadratic space in $d$, using the separator  $S' = \{v_3,v_5,v_8\}$. It would search through all combinations of values for $v_3,v_5$ and $v_8$ (time complexity $(O(|d|^|S|)$ and for each of them
determine if all of the remaining components $G_1, G_2, G_3$ and $G_4$
can be assigned a consistent value combination. For component $G_1$,
this search takes $O(|d|^2)$ time but as the result only depends on the
value of $v_3$, it can be cached so the total time complexity is $O(|d|^3)$
and space complexity $O(|d|)$. The same situation holds for components
$G_2$ and $G_3$. For $G_4$, however, all combinations of $v_3,v_5,v_8$ and $v_4$
have to be considered, and the time complexity is $O(|d|^4)$. Thus, the total time complexity is $O(|d|^4)$ time and $O(|d|)$ memory.

Intuitively, since the complexity of the best known algorithms for solving CSP depends exponentially on the treewidth, a decomposition for which a smaller bound on the treewidth of the original graph can be proven has the potential
to better preserve the minimal complexity of the original graph.
Thus, it would have been better to use the decomposition
pointed to by our Theorem.

We would pick the larger $S=\{v_3,v_4,v_5,v_8\}$ since it
allows to show a bound of $tw(G) \leq 2$ rather than $3$. When using $S$
for solving the problem, rather than searching over all combinations of values for variables in $S$, $S$ would be decomposed again into $S_1 = \{v_3\}$ and $S_2 = \{v_4,v_5,v_8\}$, where $tw(S_2) = 1$.

This shows how to solve the entire CSP in cubic time and linear space in the following steps:
\begin{enumerate}
\item first decomposition: remove $S$ and collapse the remaining graph into
vertices of $S$:
\begin{enumerate}
\item for all values of $v_3$, test whether they admit a consistent combination of $v_1$ and $v_2$ (time complexity $O(|d|^3)$, space complexity $O(|d|)$.
\item do the same for $v_5$ and $v_6,v_7$.
\item do the same for $v_8$ and $v_9,v_{10}$.
\end{enumerate}
\item second decomposition: remove $S_1$ and use search through all values of $v_3$ to:
\begin{enumerate}
\item determine and store values of $v_4$ such that there is a value of
$v_5$ that is consistent with it and the current value of $v_3$ (time complexity $O(|d|^2)$, space complexity $O(|d|)$).
\item do the same for $v_4$ and $v_8$.
\item intersect the two caches for $v_4$ and determine if any of the remaining
values is consistent with the current value of $v_3$; if yes, expand into a solution as below.
\end{enumerate}
\item Select a consistent value for $v_5$ from its respective cache, and do the same for $v_8$.
\item Use search to find combinations for $v_1$ and $v_2$ consistent with $v_3$
(time complexity $O(|d|^2)$, space complexity $O(1)$) and do the same for
$v_6,v_7,v_5$ and $v_9,v_{10},v_8$.
\end{enumerate}
The reader may verify that this algorithm requires only linear space and
cubic time in the domain size $d$, and is thus much better than the
decomposition pointed to by earlier results.

The practical lesson afforded by our Theorem is that good heuristics for
decomposing constraint satisfaction problems would look not for small
separators as current wisdom dictates, but for separators that have
few connections with each remaining component, and then apply decompositions
recursively.

As illustrated by this example, we thus believe that
Theorem~\ref{theo:little-lemma} can provide a useful supplementary
heuristic for decomposing and solving combinatorial problems using graph separators, contributing to the growing literature surveyed in \cite{BodKos09}.


\bigskip

\textbf{Acknowledgements.} The authors thank Hans Bodlaender for recommending that we highlight the
stronger statement of the Theorem as well as the Corollary.
This work was carried out when the second author was a visiting professor at the
Artificial Intelligence Laboratory (LIA), Ecole Polytechnique F\'{e}d\'{e}rale de Lausanne (EPFL).


\end{document}